\documentclass[twocolumn,showpacs,pre,amssymb]{revtex4}
\usepackage{graphicx}
\usepackage{amsmath}

\begin{document}

\title{Chaos--induced coherence in two independent food chains}

\author{Adriana Auyuanet} \email{auyuanet@fisica.edu.uy}
\affiliation{Instituto de F\'{\i}sica, Facultad de Ciencias,
  Universidad de la Rep\'ublica, Igu\'a 4225, 11400 Montevideo,
  Uruguay}

\author{Arturo C. Mart\'\i} \email{marti@fisica.edu.uy}
\affiliation{Instituto de F\'{\i}sica, Facultad de Ciencias,
  Universidad de la Rep\'ublica, Igu\'a 4225, 11400 Montevideo,
  Uruguay}

\author{Ra\'ul ~Montagne} \email{montagne@df.ufpe.br}
\affiliation{Laborat\'orio de F\'\i sica Te\'orica e Computacional,
  Departamento de F\'{\i}sica, Universidade Federal de Pernambuco,
  50670-901 Recife, PE, Brazil} 

  \date{\today}

\begin{abstract}
  Coherence evolution of two food web models can be obtained under the
  stirring effect of chaotic advection.  Each food web model sustains
  a three--level trophic system composed of interacting predators,
  consumers and vegetation. These populations compete for a common
  limiting resource in open flows with chaotic advection dynamics.
  Here we show that two species (the top--predators) of different
  colonies chaotically advected by a jet--like flow can synchronize
  their evolution even without migration interaction. The evolution is
  charaterized as a phase synchronization. The phase
  differences (determined through the Hilbert transform) of the variables
  representing those species  show a coherent evolution. 
\end{abstract}

\pacs{05.45.Xt,47.70.Fw,82.20-w}
\keywords{synchronization, chaotic advection, population dynamics}

\maketitle

\section{Introduction}
When dealing with transport processes in complex fluids flows, the
concept of turbulence comes to mind.  In bidimensional flows, it is
possible to obtain a situation where chaotic trajectories can be
generated by a simple and regular velocity field. This situation is
called chaotic advection\cite{aref02,ottino89}.  Here we show that 
chaotic advection in a oceanic jet flow can induce coherence evolution
in two chaotic systems. As an illustrative example, this simple and
robust mechanism is examined using an ocean food chain advected by
mesoscale eddies in the ocean.

In most natural habitats, numerous competing species are able to coexist, while
in general these communities are limited by only few resources (niches).  This
fact contradicts the classical theoretical and empirical studies predicting
competitive exclusion of all but the most perfectly adapted species in relation
to each limiting  factor. Recent developments in the field of chaotic advection
in hydrodynamical/environmental flows encourage us to revisit the population
dynamics of competing species in open aquatic systems. A typical model that
takes into account species interactions is a trophic web food
chain\cite{murray1,may81,nisbet1,kot01}, among them we choose a  trophic  web
food chain with a complex behavior. The complex behavior, that is, a local
disorder, is a requirement believed to be necessary for observing non trivial
collective behavior \cite{brunnet-chate}.

The article is organized as follows. In Sec.~\ref{model} we introduce the
ecological model used as well as the flow model along with the parameters 
chosen. The results of the coherent evolution of species  inmersed in a
chaotic flow are presented in Sec.~\ref{results}. The type of synchronized
regime is also investigated in that section. Our main conclusions are
summarized in Sec.~\ref{conclusions}.

\section{Models}
\label{model}
Simple models for three-species food chains exhibit a broad range of
non--equilibrium dynamics, from characteristic natural cycles to more
complex chaotic oscillations
\cite{may75,schaffer85,hasting91,murray1,hasting00,kot01}.  Two
chaotically oscillating food web models coupled diffusively may also
synchronize \cite{blasius99,blasius00}. This synchronization phenomenon
in coupled chaotic systems have been extensively studied
\cite{rosenblum96,pikovsky01}.  Those systems can display different
degrees of synchronization, namely complete synchronization, phase
synchronization, lag synchronization, and generalized synchronization
\cite{boccaletti99a}. Synchronization by periodic external actions in
the presence of noise \cite{pikovsky01} or noise--induced
\cite{toral01a} has also attracted considerable interest. Recently the
effect of stirring of chaotic advection in an inhomogeneous oscillatory
medium was investigated \cite{zoltan03}. In  the present work we study instead,
two chaotic oscillators coupled through the chaotic advection of the flow
they are immersed in.

We use a simple model of two three--species food chains 
immersed in a meandering jet flow. The flow, which is laminar and unsteady,
produces chaotic advection. One important consequence of
chaotic advection is the exponential separation of initially nearby fluid
elements. The spatio--temporal dynamics of the two colonies of food chains
embedded in the time--dependent incompressible flow is described by
the advection-reaction equations which, in a Lagrangian representation,
take the form:
\begin{equation}
\frac{d \hat{ {\mathbf r}}}{dt}= {\mathbf V}( \hat{ {\mathbf r}},t)
\label{eq:lagran1}
\end{equation}
\begin{equation}
\frac{dU_i}{dt}= F_i (U_j,{\mathbf r}= 
\hat{ {\mathbf r}}(t)) \, , \, i,j=1,2
\end{equation}
where the second set of equations describes the dynamics of the
concentration or the amount of species $U_i=\{ u_i,v_i,w_i \}$ contained
in a fluid parcel that is being advected by the flow described by the
first equation (\ref{eq:lagran1}) \cite{emilio01a}. The flow is assumed
to be imposed externally, so that the population dynamics has no
influence on the velocity field.  At scales large enough (i.e. like
ocean currents of $\approx 100 km$) diffusion effects can be neglected
\cite{abraham98}.  The coupling between the flow transport capacity and
the population evolution appears through the spatial dependence of the
$F_i (U_j,{\mathbf r})$ functions. $F_i$ varies from point to point, in
a fluid element that moves with the flow velocity. Those functions are
evaluated at the position of the fluid element at time $t$, that is, at
${\mathbf r}(t)$.

The population dynamic represented by the function $F_i (U_j,{\mathbf r})$  is a
metacomunity  explicitly modeled by two trophic food chains.  Standard three
level "vertical" food chains evolve in every parcel of a well--mixed fluid. The
resources (i.e.\ nutrients) $u_{1,2}$ are consumed by $v_{1,2}$ (i.e.\
phytoplankton), which in turn are preyed on by top predators  $w_{1,2}$ (i.e.\
zooplankton). The coupled differential equations for the biomass of the
different species are: 
\begin{eqnarray}
\frac {du_{1}}{dt} &=& a ( u_1 - u_0(\mathbf{r})) - \alpha_1
f_1(u_1,v_1) \, , %
\label{model11} \\
\frac {dv_1}{dt} &=& - b_1 v_1 + \alpha_1 f_1(u_1,v_1) - \alpha_2
f_2(v_1,w_1)\, , \label{model12}\\
\frac {dw_1}{dt}&=&  - c( w_1 - w^*) + \alpha_2 f_2(v_1,w_1) \, ,
\label{model13}
\end{eqnarray}
\begin{eqnarray}
\frac {du_2}{dt} &=&  a ( u_2- u_0(\mathbf{r})) - \alpha_1 f_1(u_2,v_2)\, ,%
\label{model21} \\
\frac {dv_2}{dt} &=& - b_2 v_2 + \alpha_1 f_1(u_2,v_2) - \alpha_2
f_2(v_2,w_2) \, , \label{model22}\\
\frac {dw_2}{dt} &=&  - c( w_2 - w^*) + \alpha_2 f_2(v_2,w_2) \, .
\label{model23}
\end{eqnarray}
The coefficients $a$, $b_{1,2}$, $c$, represent the respective net
growth rates of each individual species in the absence of interactions
among them ($\alpha_1 = \alpha_2 = 0$). Each 3--species model, in the
absence of interactions among them, has equilibrium or steady state
populations ($u_{1,2}^*, v_{1,2}^*, w_{1,2}^*$) which are the solutions of
$du_{1,2}/dt =0$, $dv_{1,2}/dt =0$, $dw_{1,2}/dt =0 $ (respectively). A
linear stability analysis yields that the steady state ($u_{1,2}^* =
0,v_{1,2}^* = 0,w_{1,2}^* = 0$), for the chosen parameters (after
Blasius et al. \cite{blasius99}) is a saddle--node point.  We set the
origin of each of the 3--species model as the steady state (in the
absence of interactions among them and uncoupled with the flow)
$u_{1,2}^{*} =0$, $v_{1,2}^{*} = 0$, $w_{1,2}^{*} = w^{*} > 0 $ . From the
population dynamics point of view, this steady state means that the
predator $w$ is allowed to maintain a low equilibrium level even when
the prey $v$, is rare. In other words there are alternative food sources
available for the predator $w$.  The two colonies can be distinguished
by a parameter mismatch of $\Delta f \approx b_2 - b_1$. The functions
$f_i$ describe interactions among the species with strengths
$\alpha_i$. We use standard interactions of Holling type II
($f_1(u,v)=\frac { u v}{1 + k_1 u}$) to describe the competition among
species $u \mbox{ and } v$. The interaction among species $v \mbox{ and
} w$ is modeled by a Lotka--Volterra interactions ($f_2(v,w)= v w$).
Equations (\ref{model11}) and (\ref{model21}) describe the evolution
of  ($u_{1,2}$), with net  rate $a$, towards a space dependent  value,
$u_0(\bf{r})$. This term, $u_0(\bf{r})$, is 
the only explicitly non--homogeneous term, it represents a spatially
dependent resource (nutrient in a plankton model) input which could
arise naturally from a variety of processes such as localized upwelling,
river run--off, translated as a source or a sink in the flow model.

The two colonies are chaotically advected by a two--dimensional
flow. The velocity field of the flow was assumed to be time dependent,
which ensures efficient mixing. Different flows have proved to produce
good stirring effect in particles, chemical reactions and plankton (see
for example \cite{emilio03a} and references therein). To illustrate this
case we choose a flow of geophysical relevance, a jet flowing eastward
with meanders, of amplitude $B(t)$ and wavenumber $k$ in the North-South
direction with a phase velocity $c_x$ \cite{cencini98}.  The cartesian
components of the flow ${\mathbf V}=( -\partial \psi / \partial y,
\partial \psi / \partial x)$ are expressed, in nondimensional units, in
terms of the stream function $\psi$
\begin{equation}
\psi(x,y) = 1 - \tanh \frac { y - B(t) \cos k (x -c_{x} t)}
{ \left( 1 + k^2 B(t)^2 \sin^2  k (x -c_{x} t)\right)^{1/2}} \, .
\end{equation}
The meander amplitude $B(t)$ is a time-dependent oscillation,  $B(t) = B_0 +
\epsilon \cos ( \omega t + \theta)$.

This flow, representing an open flow, advects eastward most of the fluid
particles, all together with the species contained in each parcel.  The
source (or sink) of resources (nutrients) $ u_0(\bf{r})$ is localized at
the origin of coordinates, according to

$u_0(x,y) = 
\begin{cases}
  1 + A \sin (\frac{ 2 \pi x } L) \sin ( \frac{2 \pi y } L) &
\text{if $x,y \in(0,L)$},\\
0 & \text{elsewhere},
\end{cases}
$

where the amplitude $A$ is constant. 

\section{Results}
\label{results}
\begin{figure}[t]
\begin{center}
\includegraphics*[height=.78\columnwidth,width=.95\columnwidth]{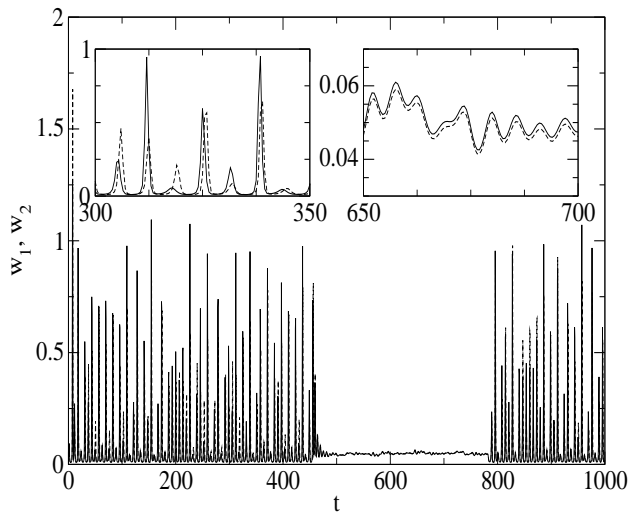}
\end{center}
\caption{Chaotic time series of top predators, $w_{1}$ and $w_2$ (dashed line) 
  in the web model.   The top images correspond to zooms of the main image. The
  image on the left shows the nonsynchronized evolution when the parcel
  goes through a region where the nutrients are homogeneous. On the
  contrary, when the parcel enters into a region where the nutrients
  are non--homogeneous (right) the top predators evolve in synchronicity.
  Parameters for the web food model are $a=1, b_1=1.1, b_2=1.055, c=10,
  k_1=0.5, \alpha_1=0.2, \alpha_2=1.0, w^*= 0.006, u_{1,2}(0)=5.0,
  v_{1,2}(0)=5.0, w_{1,2}(0)=0.0$. For the flow $B_0=1.2,
  \epsilon=0.3, \omega=0.4, c_{x}=0.12, A=0.2$.}
\label{fig1}
\end{figure}

The evolution of the colonies in the flow is integrated numerically 
according to the method proposed by Ottino \cite{ottino89} and later used
by others \cite{emilio01a,marti00a}. The two--dimensional physical
space accessible to fluid particles is subdivided into regions
characterized by different Lagrangian behaviors. The model we use,
without the spatial dependence, was shown \cite{blasius00} to have
synchronized behavior among the top predators species of the two
colonies when migration, of rate $D$ is allowed. Actually in the
absence of migration $D=0$, Blasius et al. \cite{blasius00,blasius99}
showed that the two colonies would normally be nonsynchronized. This
nonsynchronized behavior can be observed in most of the time evolution
of the top predators, as can be seen in figure \ref{fig1}. In this
figure we show the temporal evolution of the $w_{1,2}$ for a fixed
parcel in the Lagrangian point of view.

We observe in Fig.~\ref{fig1} that  the two web chains are not
synchronized at the beginning. Then, when the parcels enter into the region
where the nutrients are spatially non--homogeneous, the two colonies
start to evolve synchronously. In the left--top corner a zoom of the 
first time interval is shown. For this period of time the fluid parcel
goes through a region where the nutrients are homogeneous.  A
nonsynchronized evolution is observed, as was expected for these
parameter values \cite{blasius00}. A zoom of
the synchronous evolution can be seen in the right-top corner. This inset
corresponds to the interval when the spatial dependence, that is, the
non--homogeneous sources or sinks, advects the resources (nutrients).  This
figure shows also that the evolution of the two subsystems undergoes a
transition to another chaotic attractor, adjusting their rhythm due to the
interaction of the flow.

\begin{figure}[t]
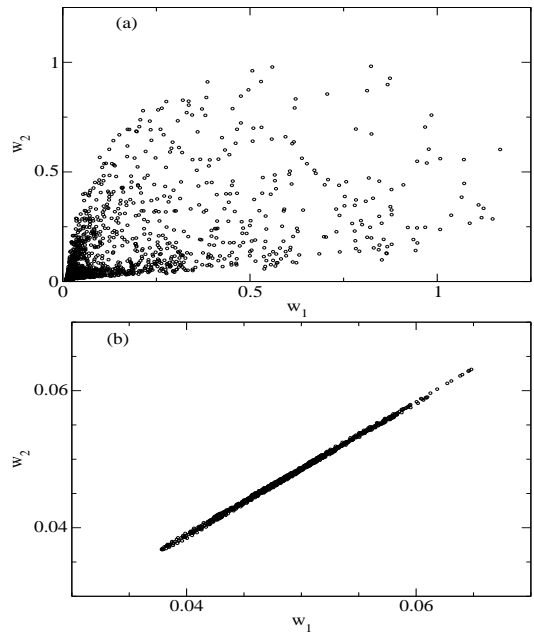

\begin{center}
\includegraphics*[height=.48\columnwidth,width=.80\columnwidth]{fig2a.eps}
\includegraphics*[height=.48\columnwidth,width=.80\columnwidth]{fig2b.eps}
\end{center}
\caption{Projections of the phase portrait on ($w_1 , w_2$)
plane. Panel (a) and (b) correspond to the cases where the nutrients
are homogeneous and inhomogeneous respectively. Notice the perfect
synchronization. Parameters for the web food model and the flow are
the same as for Fig.\ref{fig1}.}
\label{fig3}
\end{figure}

The coherent evolution of the top predators of the two colonies $w_1$
and $w_2$ can be explicitly shown plotting $w_2 \mbox{ vs }. w_1$ as in figure
\ref{fig3}. In the top panel (Fig. \ref{fig3}  a) ) $w_2 \mbox{ vs. } w_1$ is
plotted during the time where the source is not forcing the system. The clouds
of points clearly shows the uncorrelated behavior of the two variables. In Fig.
\ref{fig3} b), on the contrary, $w_2 \mbox{ and } w_1$ display a coherent
evolution. 

It is well--known that two chaotic systems could display different synchronized
regimes \cite{blasius00,boccaletti99a,pikovsky01}. Thus, we investigate the
regime of synchronization of the two colonies, as well as the influence of the
degree of mixing power of the flow.  

A closer look at the synchronized evolution reveals that this is not a complete
synchronization regime. In fact, $\overline{w_1 - w_2} \approx b_1  - b_2$
(where $\overline{...}$ means a temporal average of $...$). Another possible
scenario is that the two colonies are in a phase synchronization regime.

To describe the phase synchronization, we need to introduce
corresponding quantities. The phase of the signal (the time evolution of
the population density of one of the species) can be obtained by
different ways. We calculate the phase, using the standard construction
of the analytic signal \cite{boashash92,pikovsky01}. Complex signals are
obtained from the real signals are rewritten as a complex signals
$z_{1,2} = w_{1,2}(t) + i H[w_{1,2}(t)] = |a_{1,2}(t)|e^{i
\phi_{1,2}(t)}$, being $H[w_{1,2}(t)]$ the Hilbert transform of
$w_{1,2}(t)$ .

\begin{figure}[t]
\begin{center}
\includegraphics*[height=.60\columnwidth,width=.98\columnwidth]{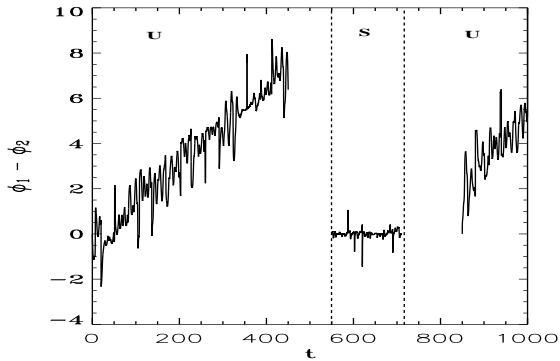}
\end{center}
\caption{Plot of phase the difference $\Delta\omega(t)= \phi_1(t) -
\phi_2(t)$ as a function of time. The regions where the phase difference
growths (meaning no synchronization) correspond to non--chaotic advection while
the regions of constant phase difference are associated with synchronized
evolution. Parameters for the web food model and the flow are the same as in
Fig. \ref{fig1}.} 
\label{fig2}
\end{figure}

 In figure \ref{fig2} we plot the relative phase difference
 $\Delta\omega(t)= \phi_1(t) - \phi_2(t)$ as a function of time. The
 method of phase estimation has several advantages and some drawbacks
 (see for a more detailed discussion \cite{boashash92} and Cap.\ 6 and
 Appendix 2 of \cite{pikovsky01}). The evolution of the point in the
 complex $(w_{1,2}(t),H[w_{1,2}(t)])$-plane rotates around two different
 centers of the two chaotic attractors. The density of population
 $w_{1,2}$ evolve in a chaotic attractor while nonsynchronized, then
 after a transient time they fell into another chaotic attractor (with
 different signal mean value) and they synchronized in phase.  After
 another transient time (where the mean value is neither of the previous
 one) they rotate once again nonsynchronously in the first chaotic
 attractor.  The transitions between the different attractors and the
 time spent in each attractor is clearly reflected in Figure
 \ref{fig2}. The first region shows the two colonies not synchronized,
 the phase difference $\Delta\omega(t)$ grows with time.  The next
 window time corresponds to the transient time when the colonies abandon
 the nonsynchronized chaotic attractor, the phase can not be estimated
 with the above procedure. Once the parcel reaches the region where the
 source (or sink) of nutrients is located, ($u_0({\bf r}) \neq 0 $), the
 phase difference between the two patches drops to a constant. The two
 species are synchronized. The two species evolve in another chaotic
 attractor with well defined mean value. Then, as the parcel leaves the
 upwelling region (or sink), there is a transient during which the
 signal leaves the attractor evolving to the nonsynchronized one. Once
 again, the phase difference in the transient time the phase can not be
 calculated. After that transient the evolution of the two species is
 once again nonsynchronized.

The flow can influence the coherence evolution of the two colonies.  Z.\
Neufeld and coworkers \cite{zoltan03} has shown that oscillators
advected chaotically by a flow can produce collective oscillations or
oscillator death by controlling the mixing capacity of the flow,
actually the stirring rate of flow. It has been shown \cite{cencini98}
that the mixing capacity of the type of flow we are using in this work
can be modified by three parameters that govern the time--dependent
oscillation of the meander amplitude, namely $B_0, \epsilon, \omega$.
We choose the values used by Cencini et al.\ \cite{cencini98} and later
by L\'opez and coworkers \cite{emilio01a} originally motivated mainly by
observations in oceans jets. These are the critical values for obtaining
"large scale chaos". It is under this situation that exchange of
particle between north-south part is more favorable (more mixing). The
parameters $B_0=1.2, \epsilon=0.3, \omega=0.4$ were chosen by Cencini et
al.\ to be greater than the critical value in order to have great power
of mixing.  Different collective behaviors of the two colonies are
expected when the mixing capacity is changed through any of the three
parameters involved, as they will be changing the mean value, the
amplitude and the stirring rate of the inhomogeneity. Different
situations may arise though, changing the parameters of the flow (as
well as the parameter of the colonies dynamic), these situations are
discussed somewhere else \cite{marti04b}.

\begin{figure}[h,t]
\begin{center}
\includegraphics*[width=.90\columnwidth]{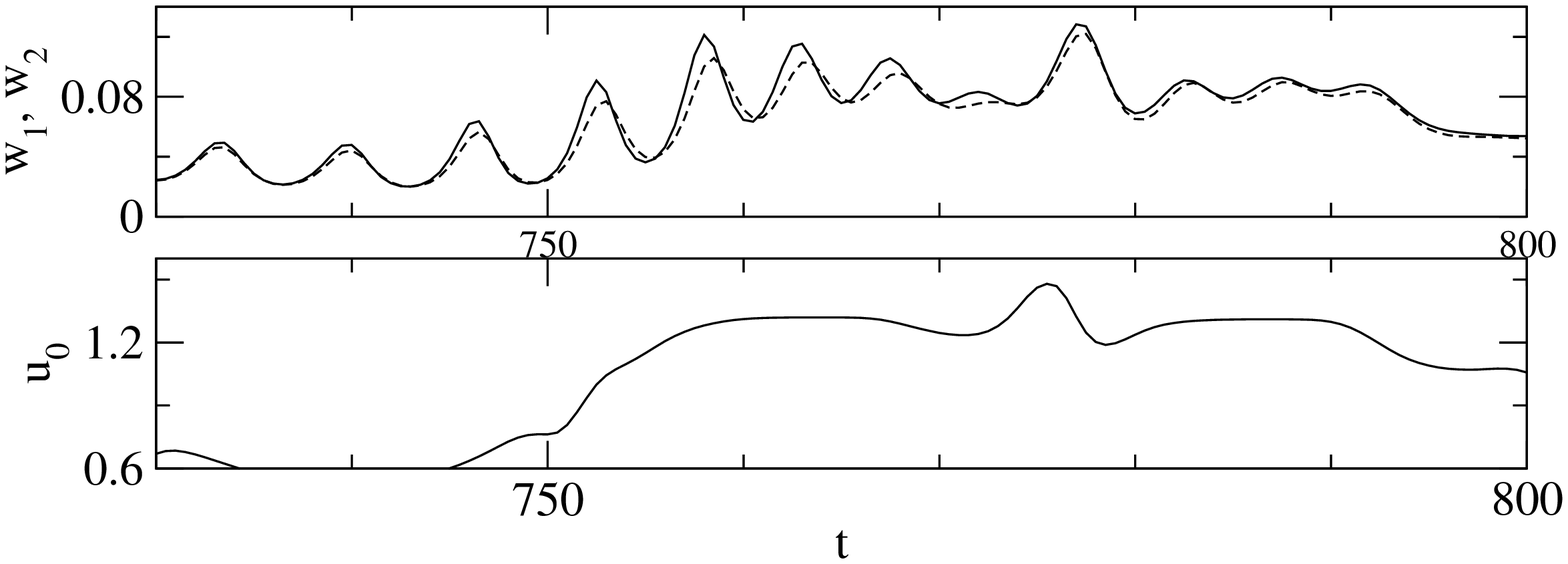}
\end{center}
\caption{The two top predators $w_{1,2}$ and the flow forcing
$u_0(\bf{r})$ as functions of time are shown.  It is shown only the
region where the nutrients are non--homogeneous. The two panels show
the independent behavior of the variables $w_{1,2}$ of the dynamical
system related to the forcing ($\omega =0.5, A=0.5$).  Parameters for
the web food model are the same as for Fig.\ref{fig1}.}  \label{fig4}
\end{figure}

We would like to remark that the two population ($w_1 , w_2$) are in
fact synchronized. That is they are not just two (chaotic) oscillators
passively following the same forcing.  It can be observed in Fig.\
~\ref{fig4} that the top-predators $w_{1,2}$ are not adiabatically
following of the forcing. Indeed for the values where the flow keeps
constant (and different from zero) the variables $w_{1,2}$ are
oscillating.

\section{Conclusions}
\label{conclusions}

In summary, we have addressed the evolution of two food web models
immersed in a flow. The chaotic advection, a mixing mechanism present
in the real ocean, was shown to induce a coherence evolution of two
species of different colonies.  In particular we have considered two
food web models of a three--species food chain, each advected by a
jet--like flow. The population model considered here represents a quite
general population dynamic, and   the main issues found here,
namely the possibility of finding coherence evolution of two species,
may be present in biological transport situations.
There still remain open questions in this issue, such as mapping in
the phase parameter space all the possible collective behavior of the
two colonies and the transition between the different attractors.
A simpler dynamical model (although less ecologically plausible) may
help to examine the influence of different flows.  Finally, we stress
that the coherence evolution of two species as a result of a mixing
property of the flow they are immersed in is a powerful process. It
has the potential to shape the distribution and abundance of aquatic
species in a current flow with important implications for ecological
dynamics in fluid flows.

\acknowledgments We acknowledge financial support from the Programa de
Desarrollo de Ciencias B\'asicas (PEDECIBA, Uruguay).  We are grateful
to E. Hern\'andez-Garc\'{\i}a for valuable suggestions.  R.M.
acknowledges useful discussions with O.\ Piro and H.\ Chat\'e.


\end{document}